\documentclass[conference]{IEEEtran}
\usepackage{cite}
\usepackage{amsmath,amssymb,amsfonts}
\usepackage{algorithmic}
\usepackage{graphicx}
\usepackage{textcomp}
\usepackage{xcolor}

\usepackage[flushleft]{threeparttable}
\usepackage{array,booktabs,makecell}
\usepackage{comment}
\usepackage{subcaption}
\usepackage{amsthm}
\usepackage{nicefrac}
\allowdisplaybreaks
\newcommand{\bsm}[1]{\boldsymbol{#1}}

\def\BibTeX{{\rm B\kern-.05em{\sc i\kern-.025em b}\kern-.08em
    T\kern-.1667em\lower.7ex\hbox{E}\kern-.125emX}}
\begin{document}

\title{Improved Syndrome-based Neural Decoder \\for Linear Block Codes}

\author{\IEEEauthorblockN{Gastón De Boni Rovella\IEEEauthorrefmark{1}\,\IEEEauthorrefmark{2}\,\IEEEauthorrefmark{3}\,\IEEEauthorrefmark{4}, Meryem Benammar\IEEEauthorrefmark{2} }
\IEEEauthorrefmark{1}TéSA Laboratory, Toulouse, France \\
\IEEEauthorrefmark{2}ISAE-SUPAERO, Université de Toulouse, France \\
\IEEEauthorrefmark{3}Centre National d'Études Spatiales, Toulouse, France \\
\IEEEauthorrefmark{4}Thales Alenia Space, Toulouse, France \\
Email: \{gaston.de-boni-rovella, meryem.benammar\}@isae-supaero.fr
}

\maketitle

\begin{abstract}
In this work, we investigate the problem of neural-based error correction decoding, and more specifically, the new so-called \emph{syndrome-based} decoding technique introduced to tackle scalability in the training phase for larger code sizes. We improve on previous works in terms of allowing full decoding of the message rather than codewords, allowing thus the application to non-systematic codes, and proving that the single-message training property is still viable. The suggested system is implemented and tested on polar codes of sizes (64,32) and (128,64), and a BCH of size (63,51), leading to a significant improvement in both Bit Error Rate (BER) and Frame Error Rate (FER), with gains between 0.3dB and 1dB for the implemented codes in the high Signal-to-Noise Ratio (SNR) regime.
\end{abstract}


\section{Introduction}
The introduction of the beyond-5G and 6G wireless technology standards in recent years has led to an increasing interest in communication tools that enable high reliability with a low processing latency. In this scenario, Machine Learning (ML) quickly gained notoriety as a powerful tool for developing fast, effective, and re-configurable solutions for most components of the digital communication chain, including error correction coding which will be of interest in this work.  

Early works in machine learning solutions for error correction coding were presented over thirty years ago \cite{Zeng_1989, Bruck_1989, Yuan_1989}, and their interest has increased dramatically ever since. Indeed, recent advances in computer science and computing power have enabled a rapid expansion in the field of Neural Networks (NN) for channel coding and decoding, producing excellent results when applied to short codes \cite{Gruber_2017, OShea_2017, Seo_2018}.

However, these solutions were promptly faced with the scaling-up problem when dealing with large code sizes. Neural-based decoders being data-driven algorithms, both their performance and generalization capability are highly dependent on the representativity of the training data. Almost optimal performances, close to Maximum A Posteriori (MAP), often require training on the entire codebook whose size is exponential in the information message length. The need to decode large codes with such intractably large training spaces gave rise to new models of scalable neural decoders \cite{Bennatan_2018_arxiv, Choukroun_2022,Nachmani_2016,Nachmani_2018,Nachmani_2021,Xu_2017}.

Currently, two main NN decoding approaches can learn to decode while training on a small subset of all possible codewords: graph-based --or model-based-- and model-free decoders.
Among graph-based solutions, Nachmani \textit{et al.} proposed a generalization of the Belief Propagation (BP) algorithm by assigning trainable weights to the edges of the Tanner Graph that characterizes a particular code \cite{Nachmani_2016}. This architecture --applied mainly to High-Density Parity Check (HDPC) codes such as BCH-- partially compensates for the effect of short cycles present in the bipartite graph. This approach quickly became dominant, and several variations were implemented \cite{Nachmani_2018, Nachmani_2021, Lugosch_2018}, including a neural BP algorithm specific to polar codes \cite{Xu_2017}, based on the BP algorithm for polar codes proposed by Arikan \cite{Arikan_2009, Arikan_2011}. All of these examples display favorable results in terms of Bit Error Rate (BER), but remain nevertheless deeply constrained to the specific structure of the code, namely the shape of its Tanner graph representation. Model-free decoders, on the opposite, usually achieve similar or better performances while keeping a more shallow neural network, and allowing for the use of more powerful ML techniques. The decoder introduced in \cite{Bennatan_2018_arxiv} displays promising BER performances while being able to incorporate a wider range of machine learning technologies \cite{Choukroun_2022}. 

However, previous works --on both approaches-- operate on a codeword level, seeking thus to minimize the errors between the sent codeword and the received one, instead of working at a message level. This can result in significant performance degradation, including invalid codewords at the output of the decoder. Furthermore, the introduction of the beyond-5G technology standard has expanded the interest in Arikan's polar codes, which are used for the control channels \cite{Arikan_2009}. Additionally, polar codes are not systematic in their original form, making Bennatan's approach not directly applicable.

For these reasons, in this work we improve the latter approach by  simultaneously fulfilling three objectives: developing a full decoder that retrieves the original message; generalizing its application to non-systematic codes; and preserving the single-codeword training property. This is accomplished while improving both bit and frame error rate measures for both polar and BCH codes.

The work is organized as follows: section \ref{sec:preliminary} presents the problem and introduces some preliminary definitions and tools. In section \ref{sec:sbnd}, we describe and justify the architecture of our solution, and propose an implementation using Recurrent Neural Networks (RNNs). Simulation results are displayed and interpreted in section \ref{sec:experiments}, and final concluding remarks are given in section \ref{sec:conclusion}, along with some future lines of work.\\ \vspace{-0.2cm}

\noindent \textit{Notation:} Roman and bold letters (e.g. $x$ and $\bsm{x}$) will be used to denote scalars and column vectors, respectively, whereas capital italic letters (e.g. $X$ and $\bsm{X}$) will represent random variables and vectors. Matrices are represented by non-italic capital letters (e.g. $\mathrm{A}$). Given $x$ a real value, $x^b$ and $x^s$ will respectively denote its hard-decision binary value (i.e. $0$ if $x>0$ and $1$ otherwise) and its corresponding Binary Phase Shift Keying (BPSK) mapping, given by $0 \rightarrow +1$, $1 \rightarrow -1$. Operations bin$(\cdot)$ and sign$(\cdot)$ are defined accordingly to perform these mappings on scalar or vector values. The Hadamard product between vectors is represented by $\odot$. Finally, $\text{P}(X\!=\!x)$ represents the probability of the event $\{ X=x \}$, i.e. the random variable $X$ taking the value $x$.

\section{Preliminary notions}\label{sec:preliminary}

\subsection{System model}\label{sec:system_model}

\begin{figure}[htbp]
    \centerline{\includegraphics[scale=0.5]{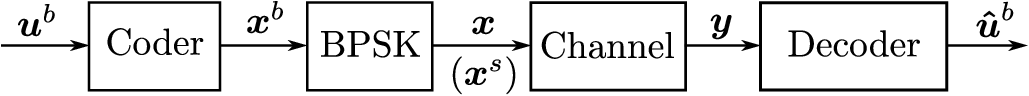}}
    \caption{General system model.}
    \label{fig:sys_model}
\end{figure}

Let us now briefly introduce the framework. A simplified schematic is given in Figure \ref{fig:sys_model}. Let $\bsm{u}^b \in \{0,1\}^k$ denote the $k$-bit message to be transmitted, and $\bsm{x}^b \in \{0,1\}^n$ its associated $n$-bit codeword through a linear code $\mathcal{C}$. This codeword is mapped to a BPSK vector $\bsm{x}\! =\! \bsm{x}^s$, which is transmitted through a symmetric binary-input Additive White Gaussian Noise (AWGN) channel.  The received signal $\bsm{y}$ is used as input to the decoder to give an estimate $\bsm{\hat{u}}^b$ of the message $\bsm{u}^b$.\vspace{0.1cm}

\subsection{Noise model}
In the traditional AWGN channel model, the received random signal is expressed as follows:
\begin{equation}
    \bsm{Y} = \bsm{X} + \bsm{N},
\end{equation}
where $\bsm{X}$ is a random vector of size $n$ that represents the BPSK modulated codeword and $\bsm{N}=(N_1, N_2, ..., N_n)$,  such that $\{N_i\}_{1\leq i \leq n}$ are independent and identically distributed (iid) random variables distributed as $\mathcal{N}(0,\sigma^2)$. In this scenario, the following holds for all $i=\{1,...,n\}$:
\begin{IEEEeqnarray}{rCl}
    \text{P}(Y_i^s \neq X_i) &=& \text{P}(Y_i^s = 1 | X_i=-1)\text{P}(X_i =-1) \nonumber \\ 
    && + \text{P}(Y_i^s = -1 | X_i=1)\text{P}(X_i =1)  \\ 
    &=& \text{P}(N_i>1)\frac{1}{2} + \text{P}(N_i<-1)\frac{1}{2}  \\ 
    &=& \text{P}(N_i>1) = \text{P}(N_i<-1). \label{eq:ni>1}
\end{IEEEeqnarray}

In this work, in order to motivate the preprocessing of the decoder input $ \bsm{y}$, we need to introduce an equivalent multiplicative formulation of the AWGN channel, which we define following \cite[Lemma 1]{Richardson_2001} by
\begin{equation}\label{eq:noise_model}
    \bsm{Y} = \bsm{X} \odot \bsm{Z},
\end{equation}
where $\bsm{X}$ and $\bsm{Y}$ designate the channel input and output vectors, and $\bsm{Z}$ is a random noise that verifies, $ \forall i=\{1,...,n\}$:
\pagebreak
\begin{IEEEeqnarray}{rCl}
    \text{P}(Z_i=z_i) &=& \text{P}(Y_i=z_i | X_i=1) \\
    &=& \text{P}(Y_i=z_ix_i | X_i=1) \\
    &=& \text{P}(Y_i=y_i | X_i=1). 
\end{IEEEeqnarray}
Therefore, $Z_i \sim \mathcal{N}(1,\sigma^2) \; $ for all $i=\{1,...,n\}$. As a direct consequence of this multiplicative model for the noise, the probability of error simplifies to:
\begin{IEEEeqnarray}{rCl}
    \text{P}(Y_i^s \neq X_i) &=& \text{P}(Z_i<0). \label{eq:zi<0}
\end{IEEEeqnarray}
It is easy to observe that the probability of a 0-centered white noise being greater than $1$  is the same as a $1$-centered white noise of the same power being smaller than $0$, i.e., \eqref{eq:ni>1} and \eqref{eq:zi<0} are equal for a same given variance $\sigma^2$.

\subsection{Polar code PC matrix and pseudo-inverse}
In the following, we introduce some results relative to polar codes that will be used later in the work. Let $\mathrm{P}_n=\mathrm{F}^{\otimes \text{log}_2n}$, where $\mathrm{F}^{\otimes k}$ represents the $k$th Kronecker power of $\mathrm{F}$, and $\mathrm{F}$ is Arikan's kernel given by
\begin{equation}
    \mathrm{F} \triangleq
    \begin{bmatrix}
        1&0\\1&1
    \end{bmatrix}.
\end{equation}
Let $\mathrm{G}$ be the $k\times n$ generator matrix for a polar code of size $n$ with $k$ information bits, composed of $k$ rows of the matrix $\mathrm{P}_n$, and let $\mathcal{A} \subset \{1,2,...,n\}$ denote the indices of these rows. Finally, let $\mathrm{V}$ represent the identity submatrix of size $k\times n$, consisting of the $k$ rows of the $n \times n$ identity matrix $\mathrm{I}_n$ with indices in $\mathcal{A}$, such that $\mathrm{G} = \mathrm{VP}_n$. Considering that the matrix $\mathrm{P}_n$ is an involutory matrix, i.e. $\mathrm{P}_n^{-1}=\mathrm{P}_n$, we introduce the following lemma:\\ \vspace{-0.2cm}

\noindent \textbf{Lemma 1.} In the previous scenario regarding a polar code $\mathcal{C}$ generated by $\mathrm{G}$, the two following statements hold:
\begin{enumerate}
    \item The matrix $\mathrm{H}$ of size $n-k \times n$ consisting of the columns of $\mathrm{P}_n$ with indices in the complement of the set $\mathcal{A}$, i.e. $\mathcal{A}^c$, is a valid Parity-Check (PC) matrix for the code defined by $\mathrm{G}$.
    \item If $\bsm{x}^b = \mathrm{G}^T\bsm{u}^b$, then $\bsm{u}^b = \mathrm{VP}_n^T\bsm{x}^b$, and thus $f(\bsm{x}^b)=\mathrm{VP}_n^T\bsm{x}^b$ yields a possible pseudo-inverse for the $(n,k)$ polar code defined by the generator matrix $\mathrm{G}$.
\end{enumerate}

\begin{proof}
The first statement is easily provable if we consider that computing the matrix product $\mathrm{GH}^T$ consists of dot products between a row and a column of $\mathrm{P}_n$ with different indices, which is $0$ because $\mathrm{P}_n\mathrm{P}_n=\mathrm{I}_n$. The second statement can be proved by expressing the generator matrix $\mathrm{G}$ as a function of $\mathrm{P}_n$ and $\mathrm{V}_n$,
\begin{IEEEeqnarray}{rCl}
     \bsm{x}^b &=& \mathrm{G}^T\bsm{u}^b =\mathrm{P}_n^T\mathrm{V}^T\bsm{u}^b,  
\end{IEEEeqnarray}
and then identifying $\mathrm{P}_n\mathrm{P}_n = \mathrm{I}_n$ and $\mathrm{VV}^T = \mathrm{I}_k$ to find the expression for the uncoded message as a function of the codeword:
\begin{IEEEeqnarray}{rCl} \label{pinv_polar}
     \bsm{u}^b &=& \mathrm{VP}_n^T\bsm{x}^b. \;\;\;  
\end{IEEEeqnarray}
This completes the proof. 
\end{proof}
As previously stated, a function that transforms every valid codeword $\bsm{x}^b$ of a code $\mathcal{C}$ into its corresponding message $\bsm{u}^b$ will be called \textit{pseudo-inverse} and will be expressed as:
\begin{equation}
    \text{pinv}(\bsm{x}^b) = \bsm{u}^b.
\end{equation}
Observe that, given a pseudo-inverse pinv$(\cdot)$, its application to an invalid codeword may yield an unpredictable result.

\section{Syndrome-based neural decoder}\label{sec:sbnd}
In this section, we first present the previous solutions that motivated our work and then introduce the proposed solution and its characteristics. 

\subsection{Previous works}
Let $\mathcal{C}$ be a linear code generated by a matrix $\mathrm{G}$, and let $\mathrm{H}$ be a PC matrix for the code $\mathcal{C}$. It was proven in \cite{Bennatan_2018_arxiv} that knowledge of the syndrome $\mathrm{H}\bsm{y}^b$ and the module of the channel output $|\bsm{y}|$ is sufficient to estimate if a position $i$ has suffered a \textit{bit-flipping} error, that is, $x^b_i \neq y^b_i$ (or equivalently, $x^s_i \neq y^s_i$), without incurring in any intrinsic loss of optimality. Building on this, two main solutions were proposed that achieve the best results among the codes simulated: the syndrome-based estimator in \cite{Bennatan_2018_arxiv} and the error correction transformer in \cite{Choukroun_2022}. However, two main issues arise:

\begin{enumerate}
    \item Most of the work in the literature focuses mostly on bit-wise codeword estimation, where the outputs are in no way restricted to a valid codeword \cite{Bennatan_2018_arxiv, Nachmani_2018, Nachmani_2021, Choukroun_2022}.
    \item There is no distinction between the information and parity bits when training the system to learn to decode. Boosting the correction capabilities of the information bits to the detriment of the parity bits could provide an increase in information BER and Frame Error Rate (FER) performance.
\end{enumerate}

The decoder in \cite{Lugosch_2018} tackled the first point, with rather modest improvements in overall performance. Our system will take it one step further by directly estimating the message $\bsm{u}^b$, and therefore removing the concept of \textit{invalid codewords} altogether. Consequently, the proposed solution will ensure full decoding of the received signal $\bsm{y}$ into a message $\bsm{\hat{u}}^b$, minimizing the error message-wise instead of codeword-wise.

\subsection{Proposed decoder}\label{sec:decoder}

In this section, we introduce a Syndrome-Based Neural Decoder (SBND) architecture that extends the work of Bennatan \textit{et al}. \cite{Bennatan_2018_arxiv} to a full decoder of the message $\bsm{u}^b$. Hence, we define a new measure of error that assesses messages instead of codewords. Let $\bsm{\tilde{u}}^b$ denote a \textit{noisy} message defined by:
\begin{equation}
    \bsm{\tilde{u}}^b \triangleq \text{pinv}(\bsm{y}^b),
\end{equation}

\noindent where the operation pinv$(\cdot)$ is a pseudo-inverse for the code $\mathcal{C}$, which can be defined as \eqref{pinv_polar} for non-systematic polar codes or vector slicing for systematic codes. Let $\bsm{w}^b$ represent the error vector between the original message $\bsm{u}^b$ and $\tilde{\boldsymbol{u}}^b$:
\begin{equation}
    \bsm{w}^b \triangleq \bsm{\tilde{u}}^b \oplus \bsm{u}^b,
\end{equation}
or equivalently, in its \emph{sign} form,
\begin{equation}
    \bsm{w}^s \triangleq \bsm{\tilde{u}}^s \bsm{u}^s.
\end{equation}
Observe that the so-called noisy message is not actually a signal we receive, but rather the output of a hard-decision decoder that thresholds the vector $\bsm{y}$ to obtain $\bsm{y}^b$ and then inverts it through the function pinv$(\cdot)$.\\ \vspace{-0.5cm}
\begin{figure}[htbp]
    \centerline{\includegraphics[width=\linewidth]{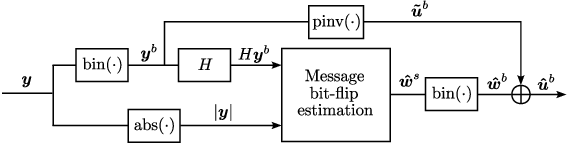}}
    \caption{SBND architecture.}
    \label{fig:sbd_architecture}
\end{figure}

Figure \ref{fig:sbd_architecture} shows the general architecture of the SBND. Essentially, the estimator uses the same inputs as \cite{Bennatan_2018_arxiv}, but will now output a vector that indicates the positions of bit-flips in the artificial vector $\bsm{\tilde{u}}^b$, which will be corrected in the final stage to obtain the estimate $\bsm{\hat{u}}^b$.\\
\vspace{-0.3cm}

\noindent \textbf{Theorem 1.} Considering the previous structure for estimating the original message $\bsm{u}^b$,  the following equation holds:
\begin{IEEEeqnarray}{l}
    \text{P}(\bsm{U}^b =\bsm{u}^b | \bsm{Y} = \bsm{y}) =  \nonumber \\
    \text{P}(\bsm{W}^s = \bsm{u}^s \bsm{\tilde{u}}^s | \; |\bsm{Z}| = |\bsm{y}| , \mathrm{H}\bsm{Z}^b = \mathrm{H}\bsm{y}^b).
\end{IEEEeqnarray}
\textit{Proof.} See Appendix.

This indicates that knowing $\bsm{y}$ and computing the probability distribution of $\bsm{U}^b$ is equivalent to knowing $\mathrm{H}\bsm{y}^b$ and $|\bsm{y}|$ and computing the probability distribution of the random variable $\bsm{W}^s$, which multiplied by the artificial variable $\bsm{\tilde{U}}^s$ yields the estimate $\bsm{\hat{U}}^s$. This extends the previous results \cite{Bennatan_2018_arxiv, Nachmani_2018, Nachmani_2021} to a full decoder architecture, where the output is the estimate of the original message $\bsm{u}^b$, and is independent of the generator matrix --and particularly, whether it is systematic or not. 

Finally, observe that Theorem 1 implies that the posterior distribution $ \text{P}(\bsm{U}^b =\bsm{u}^b | \bsm{Y} \! = \! \bsm{y})$ depends only on the noise random variable $\bsm{Z}$ and is invariant with respect to the transmitted codeword. This enables single-codeword training, as long as the noise remains random throughout the learning process.


\subsection{Noise estimation using RNN}\label{sec:noise_estimation_rnn}

\begin{figure}[htbp]
    \centerline{\includegraphics[width=0.85\linewidth]{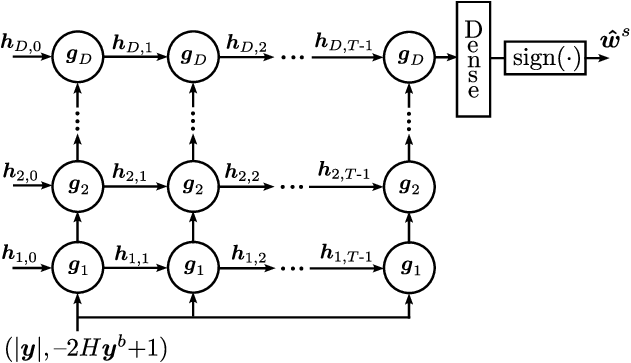}}
    \caption{A RNN implementation of the message bit-flip estimator of Figure \ref{fig:sbd_architecture}. $\bsm{h}_{i,t}$ represents the state of the $i$th GRU cell $\bsm{g}_i$ at the time step $t$.}\vspace{-0.5cm}
    \label{fig:rnn_architecture}
\end{figure}

Let us now introduce a possible implementation for the estimator of the bit-flip vector $\bsm{\hat{w}}^s$ using RNNs. Other neural estimators could also be considered in future works to further reduce the number of trainable parameters. The basic architecture is depicted in Figure \ref{fig:rnn_architecture}, where $D$ recurrent layers are stacked on top of each other and perform $T$ time steps before producing an output $\bsm{\hat{w}}^s$. Each cell $\bsm{g}_i, \; \forall i=\{1,...,D\}$ is composed of several Gated Recurrent Units (GRU) \cite{Cho_2014}, and $\bsm{h}_{i,t}$ designates the state vector of the $i$th GRU cell at time $t$. 

The input is the concatenated vector $(|\bsm{y}|,-2\mathrm{H}\bsm{y}^b+1)$ of size $2n\! -\!k$, where the BPSK mapping has been applied to the syndrome for symmetry. Each cell has $M(2n-k)$ units, where $M$ is a scaling factor hyperparameter. A final dense layer provides a vector output of $k$ elements, confined to the interval $[-1,1]$ via the tanh$(\cdot)$ activation function. A sign operation is added at the output to retrieve the actual sign vector $\bsm{\hat{w}}^s$.

\subsection{Standardization of the PC matrix}
Let $\mathrm{S}$ denote a $n_1\times n_2$ matrix of full row rank, with $n_1 \leq n_2$. $\mathrm{S}$ is said to be in its standard form when it can be expressed as follows:\vspace{-0.1cm}
\begin{equation}
    \mathrm{S} = (\mathrm{I}_{n_1} \; | \; \mathrm{S}_r),
\end{equation}
only by swapping columns and where $\mathrm{S}_r$ designates a $n_1 \times n_2-n_1$ matrix. It was observed during training that using a PC matrix in its standard form resulted in a more smoothly-decaying loss and better accuracy of estimation. For this reason, \cite[Theorem 5.5]{Hill_1990} was applied on the PC matrix $\mathrm{H}$ in order to make $\mathrm{I}_{n_1}$ a submatrix of $\mathrm{H}$. Columns swaps are omitted so as to preserve the same row-generated vector space. Further research on the influence of $\mathrm{H}$ in the SBND is being carried out. Finally, let us observe that only the PC matrix is modified, and with it the syndrome probability distribution. The code $\mathcal{C}$ and its associated matrix $\mathrm{G}$ remain unaltered.

\section{Experiments}\label{sec:experiments}

\subsection{Training and testing}

In this section, we implement, train, and evaluate the SBND previously proposed. We will use Google's TensorFlow library \cite{MartinAbadi_2015} and the Keras API \cite{Chollet_2015}. Training data are generated \textit{on the fly} in batch sizes of size $4096$, using the all-$1$ message $\bsm{u}^b = (1,1,...,1)^T$ along with an AWGN with a variance set to meet a normalized signal-to-noise ratio $E_b/N_0 = 3$dB. The network is composed of $D=5$ layers, i.e., $5$ GRU cells stacked on top of each other, and each unit performs $T=5$ time steps. Training is carried out using the Adam optimizer \cite{Kingma_2014} with a learning rate of $10^{-3}$. Table \ref{tab:parameters} shows a summary of the hyperparameters used for training. Let us observe that the same values were used for all the decoders implemented, showing robustness to the choice of parameters and suggesting the possibility of fine-tuning for further performance improvements.

\begin{table}
    \centering 
    \vspace{0.3cm}
    \begin{threeparttable}
        \begin{tabular}{ccc}
        Parameter  & Symbol & Value\\
        \midrule\midrule
        Scaling factor  &   $M$    &   $6$  \\
        \midrule
        Time steps  &   $T$    &   $5$  \\
        \midrule
        Network depth  &   $D$    &   $5$  \\
        \midrule
        Batch size &  - & $2^{12}$ \\ 
        \midrule
        Training $E_b/N_0$ & - & $3$dB \\ 
        \midrule
        Learning rate    &  $\mu$    & $10^{-3}$  \\
        \midrule\midrule 
        \end{tabular}
    \end{threeparttable}
    \caption{Model and training parameters}
    \vspace{-0.1cm}
\label{tab:parameters}
\end{table}
 
The loss function $L$ used in training is a scaled binary cross-entropy, modified to admit input values in the range $[-1,1]$, 
\begin{equation}
    L(\bsm{w}_s,\bsm{\hat{w}}_s) = \mathcal{H}\left (\frac{1-\bsm{w}_s}{2}, \frac{1-\bsm{\hat{w}}_s}{2} \right),
\end{equation}
where, for two vectors $\bsm{a} = \{a_i\}_{1\leq i \leq k}$ and  $\bsm{\hat{a}} = \{\hat{a}_i\}_{1\leq i \leq k}$,
\begin{equation}
    \mathcal{H}(\bsm{a},\bsm{\hat{a}}) = \sum_{i=1}^{k} \biggl( a_i\text{log}(\hat{a}_i) + (1-a_i)\text{log}(1-\hat{a}_i) \biggr) .
\end{equation}

To estimate the BER and FER for a given $E_b/N_0$, a Monte Carlo simulation is carried out, with a stopping criterion of $300$ frame errors and with a minimum of $10^4$ frames sent. The SBND is employed on two polar codes of rate $1/2$, with $n\! \in \!\{128,64\}$ and $k\! \in  \!\{64,32\}$, and a BCH code of $n\!=\!63$ and $k\!=\!51$\footnote{All of the parity-check matrices are taken from RPTU's website: https://rptu.de/en/channel-codes.}. For each of the three codes, our solution is compared with the best among the previously introduced model-free solutions for those particular codes \cite{Bennatan_2018_arxiv, Choukroun_2022} which we reproduced ourselves. 

BER and FER are evaluated in a message-wise sense, as opposed to the codeword-wise approach of previous works. Given that \cite{Bennatan_2018_arxiv} and \cite{Choukroun_2022} estimate the codeword $\bsm{x}^b$, we perform a pseudo-inverse to retrieve the estimated message $\bsm{\hat{u}}^b$ for non-systematic codes. For systematic codes, this comes down to extracting the systematic bits of the estimated codeword.

\subsection{Simulation results and complexity}

\begin{figure*}
     \centering
     \begin{subfigure}[b]{0.49\textwidth}
        \centering
        \includegraphics[width=\linewidth]{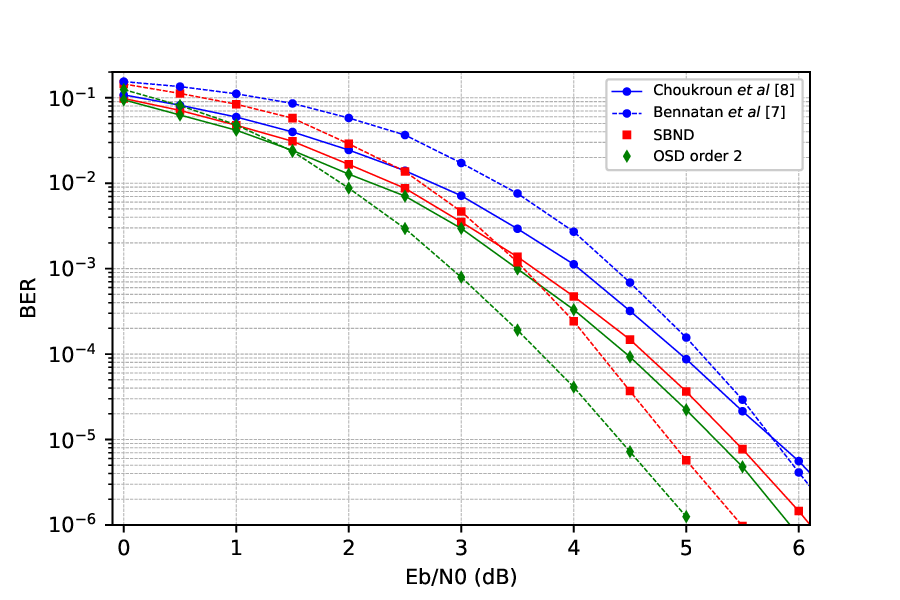}
        \label{fig:BER_polar}
     \end{subfigure}
     \hfill
     \begin{subfigure}[b]{0.49\textwidth}
     \centering
        \includegraphics[width=\linewidth]{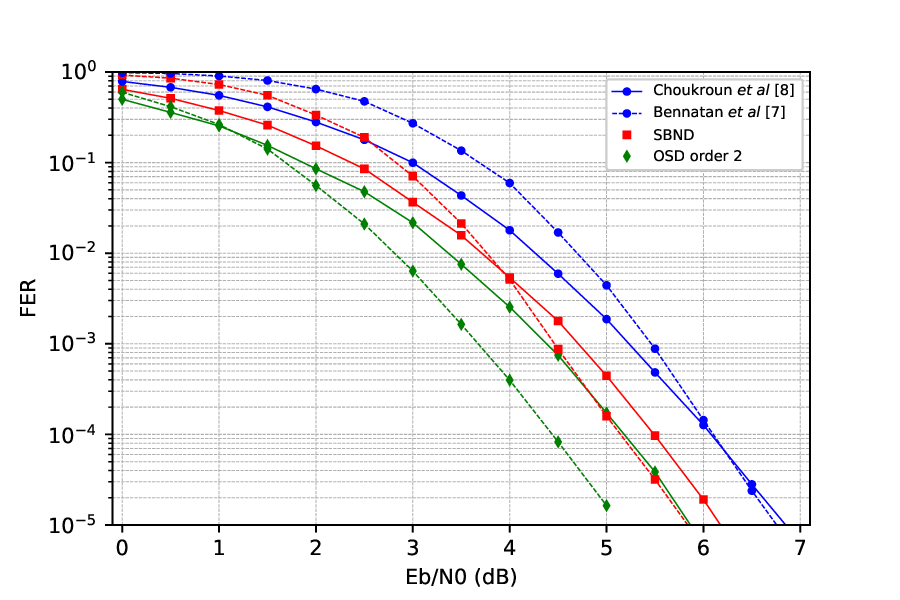}
        \label{fig:FER_polar}
     \end{subfigure}
        \vspace{-0.5cm}
        \caption{Error rate studies for two polar codes of block lengths $64$ and $128$ and code rate $1/2$. The continuous lines and the dotted lines represent the $(64,32)$ and $(128,64)$ polar codes, respectively.}\vspace{-0.4cm}
        \label{fig:polar_codes}
\end{figure*}

\begin{figure*}
     \centering
     \begin{subfigure}[b]{0.49\textwidth}
        \centering
        \includegraphics[width=\linewidth]{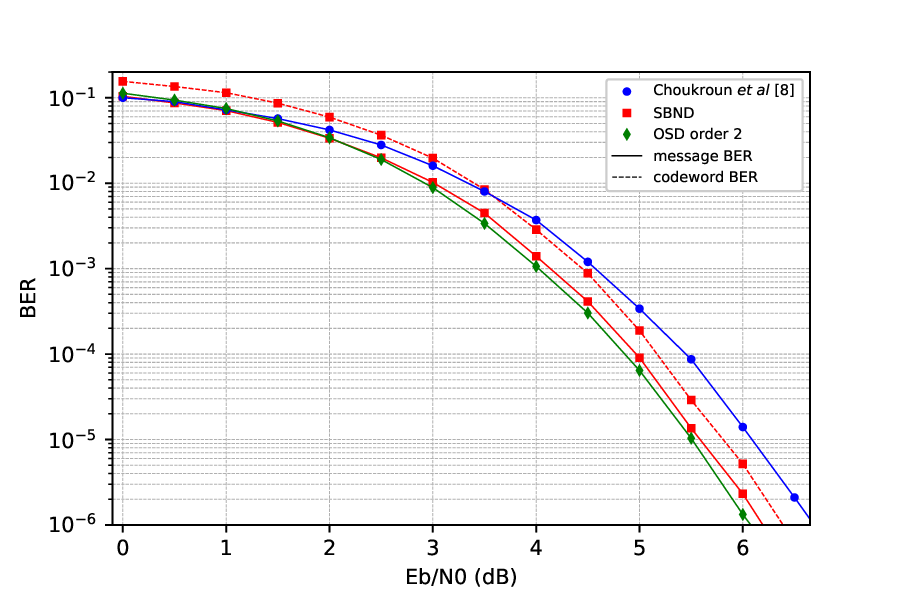}
        \label{fig:BER_bch}
     \end{subfigure}
     \hfill
     \begin{subfigure}[b]{0.49\textwidth}
     \centering
        \includegraphics[width=\linewidth]{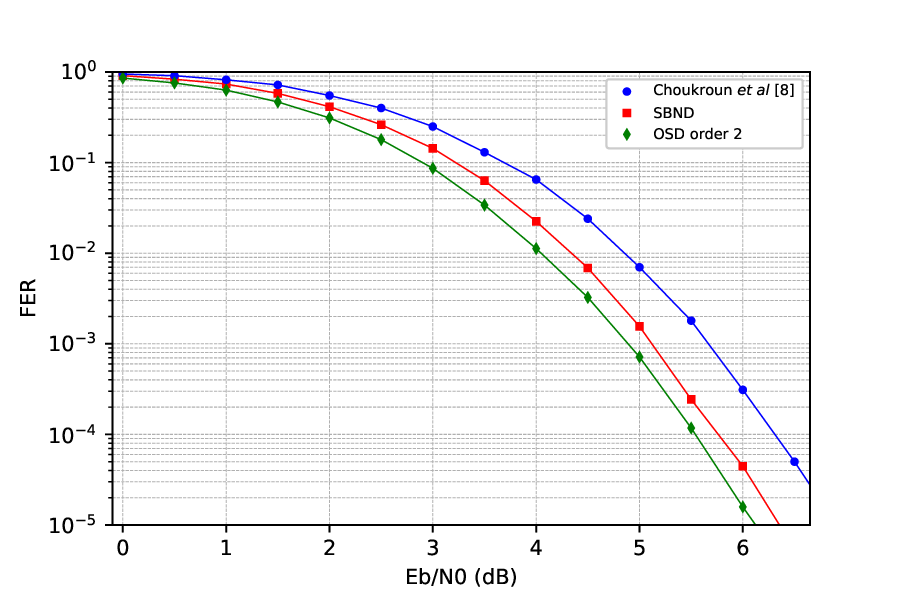}
        \label{fig:FER_bch}
     \end{subfigure}
        \vspace{-0.5cm}
        \caption{Error rate studies for a ($63,51$) BCH code. Continuous lines represent message-to-message error rates and dotted lines depict codeword-to-codeword error rates. Codeword and message BER are the same for Choukroun \textit{et al} \cite{Choukroun_2022}.}\vspace{-0.2cm}
        \label{fig:bch_codes}
\end{figure*}

Figure \ref{fig:polar_codes} shows the BER and FER performances for two polar codes of codeword size $128$ and $64$ and code rate $1/2$, respectively, through an AWGN channel. As expected, the focus on information bits results in an improvement in information BER for both polar codes. However, the most valuable gain appears in the FER curves: directly estimating messages and hence removing invalid codewords as a possible output leads to a major performance enhancement. Our system is very close to the performance of the Order Statistics Decoder (OSD)\cite{Fossorier_1995} of order 2 for the polar code of size $64$ and greatly narrows down the gap for the code of size $128$, and this for only a small fraction of the processing time required for the OSD. The performance gap for the $(128,64)$ polar code illustrates, however, the predicament of exploring a highly dimensional syndrome space, which is one of the main limitations to be tackled in the future.

Regarding complexity for the $(64,32)$ polar code decoder, the SBND has $6.3$M weights for a $5$-layer deep RNN whereas \cite{Choukroun_2022} has only $1.2$M weights but spread through a $15$-layer network, resulting in a larger decoding latency. For the $(128,64)$ polar code, the SBND and \cite{Bennatan_2018_arxiv} are very similar in number of weights, with approximately $25.5$M each. \cite{Choukroun_2022} reduced the number of weights by a factor of $10$ but with $1$dB to $1.5$dB loss in BER performance and a $23$-layer deep network.

In Figure \ref{fig:bch_codes}, our method is applied to the BCH($63,51$) and compared again with the attention-based system in \cite{Choukroun_2022}. Here we also display the codeword-to-codeword BER along with the information BER. Observe that, in the codeword sense, our solution is not better than the transformer for all the $E_b/N_0$ ranges but surpasses it in the low-noise regime. Nevertheless, on message-to-message BER, the SBND represents an improvement of approximately $0.5$dB with respect to \cite{Choukroun_2022}. The same goes for the FER, paying particular attention to the considerable proximity to the OSD of order 2. 

For this code, the SBND includes $3.9$M parameters in $5$ layers, whereas \cite{Choukroun_2022} has $1.2$M parameters over a total of $15$ layers. A deeper analysis on complexity and latency is left for a subsequent specific study.

\section{Conclusion}\label{sec:conclusion}
In this work, we have presented an extension of the system introduced by Bennatan \textit{et al.} in \cite{Bennatan_2018_arxiv}. This solution was inspired by the idea of restricting the system's output exclusively to valid codewords and finally developed into a full decoder that produces an estimation of the message.

Theorem 1 provided the mathematical basis and the possibility of training the system with only one codeword and random noise. We then presented a possible implementation of the message  bit-flip estimator using RNNs and commented on the standardization of the PC matrix for a smoother training process. Finally, the simulation results showed considerable decoding improvements with respect to the solutions in \cite{Bennatan_2018_arxiv} and \cite{Choukroun_2022}, of between $0.3$dB and $1$dB in BER and between $0.5$dB and $1$dB in FER.

For future works, the main challenge remains the scalability of the system to larger codes. Machine learning solutions seem to be limited by both the codeword and the syndrome: poorer performances on larger codes with lower code rates indicate a difficulty in properly learning over high dimensional spaces with growing sizes of input and output. Notwithstanding, the experimental results of this work suggest a very powerful solution for medium block lengths, and more advanced machine learning techniques --e.g. inspired by computer vision and natural language processing-- could allow for more precise learning over a high dimensional space without resorting to ever-growing deep neural networks.

\section*{Acknowledgements}
The present work was funded by the Centre National d'Études Spatiales (CNES) and Thales Alenia Space (TAS) and the authors would like to thank Hugo Meric from CNES and Tarik Benaddi from TAS for their helpful comments and suggestions regarding this work.

\section*{Appendix}\label{sec:appendix}
This section will provide proof for Theorem 1 of section \ref{sec:decoder}. Let us start by recalling the two claims of Lemma 1 in \cite{Bennatan_2018_arxiv}, regarding the framework of section \ref{sec:system_model}:
\begin{enumerate}
    \item There exists a matrix $\mathrm{A}$ with dimensions $k \times n$ such that $\mathrm{A}\bsm{x}^b = \bsm{u}^b$ for all possible $\bsm{u}^b \in \{0,1\}^k$ and its corresponding $\bsm{x}^b$ through the code $\mathcal{C}$, that is, $f(\bsm{x}^b)\! =\! \mathrm{A}\bsm{x}^b$ is a pseudo-inverse for $\mathcal{C}$.
    \item Given a matrix $\mathrm{B}=[\mathrm{H}^T, \mathrm{A}^T]$, then $\mathrm{B}$ has full column rank and is thus injective.
\end{enumerate}
To these results, we add and prove the following lemma:\vspace{0.2cm}

\noindent \textbf{Lemma 2.} For the random vectors $\bsm{U}^s$, $\bsm{Y}$ and $\bsm{\tilde{U}}^s$ defined as in sections \ref{sec:system_model} and \ref{sec:decoder}, the events $\mathcal{E}_1=\{\bsm{U}^s=\bsm{u}^s|\bsm{Y}=\bsm{y}\}$ and $\mathcal{E}_2=\{\bsm{U}^s\bsm{\tilde{U}}^s =\bsm{u}^s \bsm{\tilde{u}}^s|\bsm{Y}=\bsm{y}\}$ are equivalent.\vspace{0.1cm}


\begin{proof}
    Considering that $\bsm{Y}=\bsm{y}$ and that $\bsm{\tilde{u}}^s$ is a deterministic function of $\bsm{y}$, it is trivial that $\mathcal{E}_1$ implies $\mathcal{E}_2$. Additionally, since $\bsm{\tilde{u}}^s\bsm{\tilde{u}}^s=1$, then $\bsm{U}^s\bsm{\tilde{U}}^s\bsm{\tilde{U}}^s=\bsm{U}^s$. Therefore, the event $\mathcal{E}_2$ allows to unequivocally restore $\bsm{U}^s$, and thus implying $\mathcal{E}_1$.
\end{proof}

\noindent With these results, we can proceed to prove Theorem 1.
\begin{IEEEeqnarray}{l}
\text{P}(\bsm{U}^b =\bsm{u}^b | \bsm{Y} = \bsm{y}) 
=\text{P}(\bsm{U}^s =\bsm{u}^s | \bsm{Y} = \bsm{y}) \nonumber \\
\stackrel{(a)}{=}\text{P}(\bsm{U}^s\bsm{\tilde{U}}^s\!=\!\bsm{u}^s \bsm{\tilde{u}}^s | \; \bsm{Y} \!=\! \bsm{y}) \nonumber \\
\stackrel{(b)}{=}\text{P}(\bsm{W}^s \!=\!\bsm{u}^s \bsm{\tilde{u}}^s | \; |\bsm{Z}| \!=\! |\bsm{y}|, \bsm{Y}^b \!=\! \bsm{y}^b) \nonumber \\
\stackrel{(c)}{\!=\!}\text{P}(\bsm{W}^s \!=\!\bsm{u}^s \bsm{\tilde{u}}^s | \; |\bsm{Z}| \!=\! |\bsm{y}|, \mathrm{B}\bsm{Y}^b \!=\! \mathrm{B}\bsm{y}^b) \nonumber \\
\stackrel{(d)}{=}\text{P}(\bsm{W}^s \!=\!\bsm{u}^s \bsm{\tilde{u}}^s | \; |\bsm{Z}| \!=\! |\bsm{y}|, \mathrm{H}\bsm{Y}^b \!=\! \mathrm{H}\bsm{y}^b, \mathrm{A}\bsm{Y}^b \!=\! \mathrm{A}\bsm{y}^b) \nonumber \\
\stackrel{(e)}{=}\text{P}(\bsm{W}^s \!=\!\bsm{u}^s \bsm{\tilde{u}}^s | \; |\bsm{Z}| \!=\! |\bsm{y}|, \mathrm{H}\bsm{Z}^b \!=\! \mathrm{H}\bsm{y}^b, \mathrm{A}\bsm{X}^b \oplus A\bsm{Z}^b \!=\! \mathrm{A}\bsm{y}^b) \nonumber \\
\stackrel{(f)}{=}\text{P}(\bsm{W}^s  \!= \! \bsm{u}^s \bsm{\tilde{u}}^s | \; |\bsm{Z}|  \!= \! |\bsm{y}|, \mathrm{H}\bsm{Z}^b  \!= \! \mathrm{H}\bsm{y}^b, \bsm{U}^b \oplus \mathrm{A}\bsm{Z}^b  \!= \! \mathrm{A}\bsm{y}^b) \nonumber \\
\stackrel{(g)}{=}\text{P}(\bsm{W}^s  \! = \! \bsm{u}^s \bsm{\tilde{u}}^s | \; |\bsm{Z}|  \!= \! |\bsm{y}| , \mathrm{H}\bsm{Z}^b  \!= \! \mathrm{H}\bsm{y}^b).   
\end{IEEEeqnarray}

The first equation is trivial. To obtain $(a)$, we used Lemma 2. In $(b)$, we used the definition of $\bsm{W}^s$ and decomposed the variable $\bsm{Y}$ into its module and sign, where $|\bsm{Y}|=|\bsm{Z}|$ by \eqref{eq:noise_model}. In $(c)$ and $(d)$, the second claim of Lemma 1 was employed. In $(e)$, we expressed $\bsm{Y}^b$ as $\bsm{X}^b \oplus \bsm{Z}^b$, and exploited the validity of the codeword $\bsm{X}^b$:
\begin{IEEEeqnarray}{rCl}
    \mathrm{H}\bsm{Y}^b &=& \mathrm{H}(\bsm{X}^b \oplus \bsm{Z}^b)
     = \mathrm{H}\bsm{Z}^b.
\end{IEEEeqnarray}
The pseudo-inverse $\textsc{A}\bsm{X}^b = \bsm{U}^b$ was employed to obtain $(f)$. 

\noindent Finally, $(g)$ made use of the following result:
\begin{IEEEeqnarray}{rCl}
    \bsm{W}^b &=& \bsm{U}^b \oplus \mathrm{A}\bsm{Y}^b \\ \vspace{-0.45cm}
    \nonumber \\ 
    &=& \bsm{U}^b \oplus \mathrm{A}(\bsm{X}^b \oplus \bsm{Z}^b)  \\
    &=& \bsm{U}^b \oplus \bsm{U}^b \oplus \mathrm{A}\bsm{Z}^b  \\
    &=& \mathrm{A} \bsm{Z}^b \; \perp \; \bsm{U}^b \oplus \mathrm{A}\bsm{Z}^b,
 \end{IEEEeqnarray}
where $\perp$ indicates independence between two random variables and $U_i^b \sim \text{Ber}(\nicefrac{1}{2}) \; \forall i \! =\!\! \{1,...,k\}$. Given that $\bsm{U}^b \oplus \mathrm{A}\bsm{Z}^b$ is independent of $\bsm{W}^b$ --and thus of $\bsm{W}^s$--, it can be removed from the conditional probability expression.  \hspace{2cm}$\square$

\bibliographystyle{IEEEtran}
\bibliography{Paper_decoder}

\end{document}